\pdfoutput=1 
\documentclass[english]{article}
\usepackage[T1]{fontenc}
\usepackage[latin9]{inputenc}
\usepackage[letterpaper]{geometry}
\usepackage{amssymb}
\usepackage{esint}
\usepackage{babel}

\begin{document}

\title{\textbf{How happy is your web browsing? A model to quantify satisfaction
of an Internet user, searching for desired information.}}

\date{\textbf{Anirban Banerji}{*}, \textbf{Aniket Magarkar}\\
Bioinformatics Centre, University of Pune, Pune-411007, Maharashtra,
India\\
(\textbf{anirbanab@gmail.com})}
\maketitle
\begin{abstract}
We feel happy when web-browsing operations provide us with necessary
information; otherwise, we feel bitter. How to measure this happiness
(or bitterness)? How does the profile of happiness grow and decay
during the course of web-browsing? We propose a probabilistic framework
that models evolution of user satisfaction, on top of his/her continuous
frustration at not finding the required information. It is found that
the cumulative satisfaction profile of a web-searching individual
can be modeled effectively as the sum of random number of random terms,
where each term is mutually independent random variable, originating
from 'memoryless' Poisson flow. Evolution of satisfaction over the
entire time interval of user's browsing was modeled with auto-correlation
analysis. A utilitarian marker, magnitude of greater than unity of
which describe happy web-searching operations; and an empirical limit
that connects user\textquoteright{}s satisfaction with his frustration
level - are proposed too. Presence of pertinent information in the
very first page of a web-site and magnitude of the decay parameter
of user satisfaction (frustration, irritation etc.), are found to
be two key aspects that dominate web-browser's psychology. The proposed
model employed different combination of decay parameter, searching
time and number of helpful web-sites. Obtained results are found to
match the results from three real-life case-studies.\\

\end{abstract}
\textbf{1. Introduction}\\
We browse through\textbf{ }hundreds of web-pages everyday to find
information of our interest. We all know how happy we become to find
some bits of relevant information, and how irritated we feel not to
find the same in scores of pages. However, familiar as we all are
with web-searching, a model that describes the characteristics of
it in general and objective terms, eludes us. In the present work,
an attempt is made to construct a quantitative framework that describes
the evolution of satisfaction level of an Internet user, while he
(she, at any rate) is searching for a (precise) piece of information
in web. In related note, we notice that there are lot of works that
attempt to address various aspects of consumer experiences and concerns
with the adoption of Internet banking {[}1-8{]}. However, problems
addressed in these studies are either of empirical nature, particular
to mindset of web-users in certain geographical locations; or are
extremely focussed in their very nature - resulting in limited amount
of web-browsing; or both. In any case, since the essence of Internet
banking is in visiting pre-defined (already known) web-sites to undertake
pre-determined operations, these approaches find little relevance
in our effort to quantify the evolution of satisfaction profile during
any web-browsing operation. Similarly, although some other works have
attempted to address the spectrum of (inter-related) issues related
to ease (or difficulties) of web-users resorting to either online-buying
{[}9{]} or online-investing {[}10{]}, they concentrated on detailed
exposition of psychological features of web-users and not on constructing
a general mathematical framework to describe the evolution of satisfaction
of an Internet user in pursuit of a specific piece of information.
Hence, although one finds many works that explore problems that are,
at best, somewhat akin to the present one; relevance of them to the
present one can at best be categorized as tangential. Thus, methodologies
used in the aforementioned works were not employed in our work.\\
\\
Since the entire process of web-searching has an inherent probabilistic
nature, a probabilistic model is proposed here to model the problem
in hand. The present model does not attempt to describe the satisfaction
level of an Internet browser when he is searching for information
easily available; like, searching for a restaurant in a particular
part of a metropolis with certain preferred dishes. Such searching
operations hardly last for minutes and seldom continue over an appreciable
span of time. In contrast, we attempt to quantify the evolution of
Internet user's satisfaction over a long interval of time; during
which he searches, finds accidental useful links but fails to find
all the relevant information, assemble and organize the obtained bits
of information, redefines his queries - and carries on till he successfully
gathers the required pertinent chunk of information he was seeking.
An example for this kind of operation might be searching for a post-doctoral
job under a capable professor (with presentable publication record)
with a very specific (and not general) goal, where the job tenure
(with a decent salary) meets a preferred duration of time and the
preferred institutions are supposed to be in a particular geographic
area. One might think of searching for a particular piece of multidisciplinary
scientific information (typically, of comprehensive nature) from various
research articles - as another example for searching operations referred
to in this study (say, for example, an account of probabilistic constructs
that have attempted to describe protein tertiary-structure organizations,
accepting the fact that proteins are fractal objects). However, although
incisive searching operations from almost all walks of life regularly
attempt to retrieve information of similar depth, algorithms used
in the search engines, often fails to account for the complexities
of searching requirements of non-trivial searching operations. As
a result, to obtain the complete piece (or chunk) of desired information,
an Internet user attempts to overcome the frustration of not accumulating
the entire bulk with satisfaction of particular moments at recognizing
helpful information. It is the evolution of this satisfaction that
we attempt to model here in this study.\\
\\
Experience suggests that, realistically, the user (say $U$) seldom
hits upon a (magical) web-site with the entire bulk of desired information.
Instead, he gathers bits and pieces of relevant information from various
web-sites. Typically, $U$ collects the whole chunk of requisite information
as an arbitrarily large number of small bits of information. Let us
assume that a flux of tiny bits of pertinent information in the very
first page of any web-site is what captures the imagination of the
user and motivates him to read through the content of the entire site
in sufficient detail. We assume that the satisfaction level of $U$
is purely a function of the level of information he is receiving from
the web-site. In other words, we assume that the visual-layout features
of the web-sites do not contribute to the existing level of information
content of $U$. Hence, for the present study, visual designing is
not considered as an influencing parameter of the satisfaction level
of $U$. We assume further that $U$'s coming across these tiny bits
of pertinent information forms an elementary flow of intensity $\lambda$
(scalar rate parameter).\\
\\
\textbf{2. Events of user's encountering helpful web-sites, form
a Poisson flow}\\
Reasons behind the assertion of elementary flow are easy to observe
{[}11-19{]}. First\textbf{ }of all, events of $U$'s coming across
these tiny bits of pertinent information do not depend upon the choice
of a reference point. In other words, probability of taking place
of concerned events (viz. $U$'s coming across favorable web-sites)
within any small length of total time-interval, depends only on the
length of total time-interval; and not on its position on time-axis.
Hence, occurrence pattern of favorable events is stationary. Second,
since at any instance of time, $U$ reads (or glances through) the
content of one web-site (and not more than that, on the ground of
physical impossibility), aforementioned occurrence pattern can be
classified as ordinary. (Therefore, one might interpret the flow of
events of $U$'s coming across favorable web-sites as stochastic.)
Third, experience suggests that $U$'s coming across favorable web-sites
on any (arbitrarily small) time interval does not depend upon $U$'s
encountering other favorable web-sites on any other non-overlapping
interval on the time-axis. In other words, the events suffer from
{}``lack of memory''{[}17{]}. Owing to its satisfying these three
conditions, $U$'s coming across favorable web-sites can be asserted
as an elementary flow.\\
\\
Here, it is worthwhile to note that even if the search operation
is carried out with multiple tabs on the web-browser open, since U
can read (and contemplate) the content of only one web-page at a time,
multiple tabs do not make the searching operation a multi-threaded
operation. Hence, throughout the present study, the searching operation
will be considered as a single-threaded operation.\\
\\
It is realistic to assume that the very moment $U$ touches upon
a piece of information that he thinks might take him closer to the
desired piece of information he is searching for, his satisfaction
level grows. To make the calculation simpler, we assume that this
growth of satisfaction in $U$, takes place with a constant scale
of unit magnitude. However, as it generally happens in real-life,
very soon $U$ realizes that any typically encountered tiny piece
of hint of an information does not always take him closer to the set
of information he wishes to possess, but is aimed at something else
that isn't exactly related to the premise of question he is interested
in. His satisfaction level therefore starts to decay. To represent
the situation reasonably, we assume this decay function to be having
an exponential nature (rather than having a two-state (or some other)
characteristics), with a parameter $\mu$. Thus, while $\lambda$
is influenced predominantly by the content of the site; origin of
$\mu$ is complex, because of its dependence on various parameters.
Nevertheless, those tiny sources of information satisfies $U$ to
some extent and a cumulative effect of these acquired information
starts to build up in his mind. We assume that the gradual growth
of residual satisfaction of $U$ over the searching time $t$, through
scores of web-page, can be captured by a linear sum. We designate
the user-satisfaction level by $X(t)$. Here a simple model is proposed
to find the characteristics of growth of user-satisfaction upon user's
web-browsing.\\
\\
\textbf{3. Methods}\\
At this point we segment our study in two cases. In the first
case we model the case when the user gradually accumulates useful
information from various web-sites. In the second case, we model the
situation when the user encounters a series of web-sites (threaded,
quite possibly); and thereby, receives the helpful information in
bursts, rather than in trickles as the first case. Model to describe
the second situation, although related to the first one, poses certain
different challenges. Hence, they are discussed differently in case-1
and case-2, respectively.\\
\textbf{3.1: Case-1 : User receiving useful information by trickle}\\
\textbf{3.1.1 :} \textbf{The basic framework}\\
Let us assume that $U$ touches upon the pertinent information
at random moments, $T_{1},T_{2},\ldots,T_{i},\ldots,$ which forms
an elementary flow of events. The user satisfaction level at any arbitrarily
chosen moment $t$, due to his encountering any $i^{th}$ bit of information
at the moment $T_{i}$, is given by :\begin{eqnarray}
S_{i}(t) & = & 0,\;(t<T_{i})\\
 & = & e^{-\mu(t-T_{i})},\;(t\geqslant T_{i})\end{eqnarray}
\\
Hence, compositely, we can write $S_{i}(t)=1(t-T_{i})\: e^{-\mu(t-T_{i})}$
, where $1(t)$ is a unit function; $T_{i}>0,\; t>0$.\\
\\
Let us now define a random variable $\Omega$ that describes the
number of such tiny bits of information influencing the satisfaction
level of $U$. This variable, to be realistic, will be having a Poisson
distribution with parameter $\lambda t$ {[}11-19{]}.\\
\\
It is clear that the user-satisfaction level $X(t)$, in realistic
terms, should be the sum of random number of random terms :\\
\begin{equation}
X(t)=\sum_{i=1}^{\Omega}e^{-\mu(t-T_{i})}\:1(t-T_{i})\end{equation}
\\
\textbf{}\\
\textbf{3.1.2 : On the exact characterization of Poisson flow}\\
Careful examination of the situation suggests that this problem
is not straightforward. Because, here we have $\alpha$ number of
mutually independent random variables $\psi_{i}\;\left(i=1,\,2,\:\ldots\:,\,\alpha\,\right)\;$,
each of which is uniformly distributed within the interval $(0,\: T)$.
For the present problem, distribution of the number $N$ of the random
variables (points) $\psi_{i}$, occupying the interval $\;\left(k,\: l\right)\subset\left(0,\: T\right)\;$,
would have to be ascertained. It assumes importance, especially for
the present case, to ascertain the limiting distribution of the random
variable $N$, when $\:\alpha\longrightarrow\infty,\: T\longrightarrow\infty\:$
and the average number of points within the interval $\left(k,\: l\right)$
remains constant.\\
\\
We start by constructing a counter, which marks every occurrence
of the event $\:\psi_{i}\:\in\,\left(k,\: l\right)$. We assign a
random variable $\, N_{i}\:$ for this counting, so that :\\
\begin{eqnarray*}
N_{i} & = & 1\qquad if\;\:\psi_{i}\:\in\,\left(k,\: l\right)\\
 & = & 0\qquad if\;\:\psi_{i}\:\notin\,\left(k,\: l\right)\end{eqnarray*}
where, $N\:=\sum_{i=1}^{\alpha}\, N_{i}$\\
\\
Furthermore, we designate :\\
$P\left(\: N_{i}\,=1\right)\,=\: p\:=\,\frac{\left(\, l\,-\, k\,\right)}{T}\;$
and $\; P\left(\: N_{i}\,=0\right)\,=\:1-p\:=\,\frac{\left(\, T\,-\, l\,+\, k\,\right)}{T}\;$
\\
\\
One may view $\alpha$ values (viz. $\,\psi_{1}\,,\,\psi_{2}\,,\:\ldots\,,\,\psi_{\alpha}\;$
) as the results of $\alpha$ independent trails, in each of which
the event $\:\psi_{i}\:\in\,\left(k,\: l\right)$ was probable. A
realated observation reveals that random variable $N\:$ has a binomial
distribution with parameters $\alpha$ and $p$, where the mean value
is given by : $E\left(N\right)=\alpha p\;$; so that we have $\: P\left(\, N\,=r\right)\,=\: C_{\alpha}^{r}\, p^{r}\:\left(1-p\right)^{\alpha-r}\;\left(\, r\,=\,0\,,\,1\,,\,\ldots\,,\,\alpha\,\right)$\\
\\
\\
We concentrate on the case when $\:\alpha\longrightarrow\infty,\: T\longrightarrow\infty\:,\:\frac{\alpha}{T}\,=\lambda=a\; constant$.
Under such a situation, $\: p\longrightarrow0$ ; but the average
number of points occupying the interval $\left(k,\: l\right)$ is
constant; viz. $E\left(N\right)=\lambda\left(\, l\,-\, k\,\right)\;=constant$.
It is known from basic probability {[}11-19{]} that the limiting distribution
of the random variable $N$ will be a Poisson distribution with parameter
$\:\beta=\lambda\left(\, l\,-\, k\,\right)\;$; whence, $\: P\left(\, N\,=r\right)\,=\frac{\beta^{r}}{r!}\, e^{-\beta}\quad\left(\, r\,=\,0\,,\,1\,,2,\,\ldots\,,\right)$.
Thus, number of events that occupy the interval $\left(k,\: l\right)$
and that are in a stationary Poisson flow with intensity $\lambda$,
will have the same distribution.\\
\\
Importance of this conclusion can be adjudged by the fact that,
since a Poisson flow of events on any interval $(0,\, t)$ can be
represented with sufficient accuracy as a collection of points on
that interval, the coordinate of a point $\alpha_{i}\in(0,\, t)$
can be considered to be uniformly distributed on that interval being
independent of the coordinates of other points. (One may expect the
same from a non-mathematical intuitive understanding of the situation
also.) Relating to the present situation, since $U$ comes across
all these points (representing the exact instance of finding a bit
of relevant information) during the interval $(0,\, t)$, one can
now be confident of the fact that; first, this description exhaustively
represents the entire event space of favorable encounter for the user
during $(0,\, t)$, and second, different web-pages with different
approaches relating to the original query that the user encounters
at different time instances are all taken into account.\\
\\
Therefore, eq$^{n}$-3 can be re-written as :\\
\begin{equation}
X(t)=\sum_{i=1}^{\Omega}e^{-\mu(t-\alpha_{i})}\end{equation}
where the random variables $\alpha_{i}$ are independent and uniformly
distributed in the interval $(0,\, t)$.\\
\\
Since the satisfaction of $U$ is a function of interplay of $\lambda$
and $\mu$, and since practical experience suggests it to be having
a complicated yet cumulative nature, we can attempt to model the user
satisfaction as a resultant of events of favorable information gathering
and events of unfavorable decay. We designate $X_{i}(t)=e^{-\mu(t-\alpha_{i})}=e^{-\mu t}e^{\mu\alpha_{i}}$,
where $X_{i}(t)$ represents each of these tiny events. Hence we have
:\\
\begin{equation}
X(t)=\sum_{i=1}^{\Omega}X_{i}(t)=e^{-\mu t}\sum_{i=1}^{\Omega}e^{\mu\alpha_{i}}\end{equation}
where $X_{i}(t)$ are independent identically distributed random variables,
and $\Omega$, a random variable that does not depend upon $X_{i}(t)$.
Here we note that although $X(t)$ is cumulative in nature, essentially
it describes a stochastic process.\\
\\
At this moment, invoking the known formula regarding mean and
variance of the sum of a random number of random variables {[}20{]}
we arrive at :\\
$m_{x}(t)=m_{\Omega}(t)m_{x_{i}}(t)\quad$ and $\quad Var_{x}(t)=m_{\Omega}(t)Var_{x_{i}}(t)+Var_{\Omega}(t)m_{x_{i}}^{2}(t)\;$,
where $m_{x}$ and $Var_{x}$ represent mean and variance of random
variable $Z$ $\left(Z=\sum_{i=1}^{\Omega}X_{i}\right)$ respectively;
$m_{\Omega}$ and $Var_{\Omega}$ represent mean and variance of integral
random variable $\Omega$ (independent of terms of $X_{i}$) respectively.\\
\\
\\
Since the random variable $\Omega$ has a Poisson distribution
with parameter $\lambda t$, it follows that $m_{\Omega}(t)=Var_{\Omega}(t)=\lambda t$.\\
\\
Furthermore, \\
we have $m_{x_{i}}(t)=E[X_{i}(t)]=\frac{1}{t}\int_{0}^{t}e^{-\mu(t-x)}dx=\frac{1-e^{-\mu t}}{\mu t}\quad$
and $\quad E[X_{i}^{2}(t)]=\frac{1}{t}\int_{0}^{t}[e^{-\mu(t-x)}]^{2}dx=\frac{1-e^{-2\mu t}}{2\mu t}$.\\
\\
Hence, $m_{x}(t)=\lambda\frac{1-e^{-\mu t}}{\mu}\quad$ and $\quad Var_{x}(t)=\lambda tE[X_{i}^{2}(t)]=\lambda\frac{1-e^{-2\mu t}}{2\mu}$.\\
\\
It is interesting to notice that as $t\rightarrow\infty$, the
mean value and variance of the process $X(t)$ do not depend on time,
since\\
$lim_{t\rightarrow\infty}m_{x}=m_{x}=\frac{\lambda}{\mu}\quad$
and $\quad lim_{t\rightarrow\infty}Var_{x}(t)=Var_{x}=\frac{\lambda}{2\mu}$.\\
\\
This is expected purely from an intuitive perspective also. After
traversing through the web-sites for a sufficiently long time, the
user is expected to gather a finite amount of desired information.
However, since he fails to remember all of it, only a fraction of
the amassed information will be retained by him. Thus the fraction
$\frac{\lambda}{\mu}$ can be named as 'satisfaction retentivity quotient'.\\
\\
\\
\textbf{3.1.3 Correlation between user satisfaction profiles}\\
Characteristics of user satisfaction can perhaps be understood
in the best way by studying the evolution of this satisfaction over
the entire time interval of user's browsing through web-sites. Hence
we proceed to find the correlation function between user satisfaction
profiles during different instances of browsing operation. This can
be done by considering two sections of the stochastic process in question,
at the instances $t$ and $t^{\prime}$$(t^{\prime}>t)$. By virtue
of the assumption made, we can assert that the user satisfaction $X(t^{\prime})$
at the moment $t^{\prime}$, is equal to the extent of satisfaction
$X(t)$ at the moment $t$ multiplied by the exponent $e^{-\mu(t^{\prime}-t)}$,
added with the satisfaction $\Omega(t^{\prime}-t)$, which comes into
being due to user's coming across some interesting bits of information
during the time interval $(t^{\prime}-t)$. Hence $X(t^{\prime})$
is given by :\\
\begin{equation}
X(t^{\prime})=[X(t)\, e^{-\lambda(t^{\prime}-t)}+\Omega(t^{\prime}-t)]\end{equation}
\\
The stochastic processes $X(t)$ and $\Omega(t^{\prime}-t)$ are
evidently independent since they are generated due to user's interaction
with desired piece of information during different, non-overlapping
time intervals $(0,t)$ and $(t,t^{\prime})$ respectively. \\
\\
The same can be said about the centered stochastic processes $\dot{X}(t)$
and $\dot{\Omega}(t^{\prime}-t)$, where we define $\dot{X}(t)=X(t)-m_{x}(t)$
and $\dot{\Omega}(t^{\prime}-t)=\Omega(t^{\prime}-t)-m_{\Omega}(t^{\prime}-t)$
as centered random functions of the aforementioned stochastic processes.\\
\\
Hence, using eq$^{n}$-12 we have : \\
\begin{eqnarray*}
C_{x}(t,t^{\prime}) & = & E\left[\dot{X}(t)\,\dot{X}(t^{\prime})\right]\\
 & = & E\left[\dot{X}(t)\,\left\{ \dot{X}(t)\, e^{-\mu(t^{\prime}-t)}+\dot{\Omega}(t^{\prime}-t)\right\} \right]\\
 & = & E\left[\left(\dot{X}(t)\right)^{2}\right]\, e^{-\mu(t^{\prime}-t)}\quad if(t^{\prime}>t)\\
 & = & E\left[\left(\dot{X}(t^{\prime})\right)^{2}\right]\, e^{-\mu(t-t^{\prime})}\quad if(t^{\prime}<t)\end{eqnarray*}
Thus the correlation can be compositely expressed as :\\
\begin{equation}
C_{x}(t,\, t^{\prime})=Var_{x}(min(t,\, t^{\prime}))\left[1-e^{2.\alpha.min(t,t^{\prime})}\right]e^{-\mu|t^{\prime}-t|}\end{equation}
\\
\\
Let us consider the limiting behavior of the stochastic process
when $t\rightarrow\infty$, $t^{\prime}\rightarrow\infty$, but the
magnitude of their difference $\tau=t^{\prime}-t$ is finite. In this
case, $C_{x}(\tau)=Var_{x}e^{-\mu|\tau|}=\frac{\lambda}{2\alpha}e^{-\mu|\tau|}$.
\\
\\
Hence the stochastic process $X(t)$ representing user satisfaction
practically attains stationarity in every aspect when the user spends
a long time searching for some desired bulk of information. This conforms
to our experience.\\
\\
Of course the user can stumble upon a web-site where the information
regarding all of his interest are kept in one place. In such (unlikely)
case, naturally $\mu\rightarrow0$. Here the extent of user satisfaction
will be a Poisson process since every new piece of information that
the user will be encountering will exactly match with the desired
set of information he wanted to collate. Consequently, the decay in
user's interest will occur minimally. In such a case, the expressions
for $m_{x}(t)$, $Var_{x}(t)$ and $C_{x}(t,t^{\prime})$ will assume
the form :\\
\\
$lim_{t\rightarrow\infty}m_{x}(t)=lim_{\mu\rightarrow0}\lambda\frac{1-e^{-\mu t}}{\mu}=lim_{\mu\rightarrow0}Var_{x}(t)=lim_{\mu\rightarrow0}\lambda\frac{1-e^{-2\mu t}}{2\mu}=\lambda t$\\
\\
$lim_{\mu\rightarrow0}C_{x}(t,t^{\prime})=lim_{\mu\rightarrow0}\frac{\lambda}{2\mu}\left[1-e^{2\mu min(t,t^{\prime})}\right]e^{-\mu|t^{\prime}-t|}=\lambda\left[min(t,\, t^{\prime})\right]$\\
\\
\\
\textbf{3.2.: Case-2 : User receiving useful information in bursts}\\
In some other real-life cases another situation is frequently
encountered. Here certain related pieces of information from a web-site
conform to user's desired set of information and user comes across
these related pieces of information in a somewhat quantized form.
Although this case is similar to one discussed already, there are
certain subtle differences. To find the characteristics of user satisfaction
level in this situation, we assume the applicability of the assumptions
made earlier and at the same time assume further that, user's coming
across such quantum of desired information form an elementary flow
with intensity $\lambda$(discussed beforehand). The exact number
of information that constitute any i$^{th}$ quantum of desired information
is assumed to be a random variable $R_{i}$, which, keeping with the
real-life situation, is obviously independent of the number of pieces
of information that constitutes any other quantum. The random variable
$R_{i}$ has a distribution $f(R)$ with characteristics of $m_{R}$
and $var_{R}$.\\
\\
Just like the case where user was encountering the desired information
in bits and pieces(eq$^{n}$3), here too we can represent the extent
of user-satisfaction by:\\
\begin{equation}
X(t)=\sum_{i=1}^{\Omega}R_{i}e^{-\mu(t-\alpha_{i})}\end{equation}
where the random variables $\Omega$, $R_{i}$ and $\alpha_{i}$ are
mutually independent.\\
\\
Keeping with the earlier approach, we designate $X_{i}(t)=R_{i}e^{-\lambda(t-\alpha_{i})}$
and then \\
$E\left[X_{i}(t)\right]=m_{R}\frac{1-e^{-\mu t}}{\mu t}$ and
$E\left[X_{i}^{2}(t)\right]=(Var_{R}+m_{R}^{2})\frac{1-e^{-2\mu t}}{2\mu t}$\\
\\
Henceforth, $m_{x}^{quant}(t)=\lambda m_{R}\frac{1-e^{-\mu t}}{\mu}\quad$
and $\quad Var_{x}^{quant}(t)=\lambda[(Var_{R}+m_{R}^{2})\,\frac{1-e^{-2\mu t}}{2\mu}]$\\
\\
Since $m_{R}>0$, $m_{x}^{quant}$ will grow faster than $m_{x}$.
This is completely in agreement with practical experiences. Since
user comes across the desired bulk of information in a coherent quantized
form, his satisfaction grows quickly.\\
\\
This basic scheme of swiftness of knowledge gathering (obviously)
doesn't change when the user browses for a long time and at the limiting
case $t\rightarrow\infty$, we have:$\quad lim_{t\rightarrow\infty}m_{x}(t)=m_{x}=\frac{\lambda m_{R}}{\mu}\;$,
$\; lim_{t\rightarrow\infty}Var_{x}(t)=Var_{x}=\frac{\lambda(Var_{R}+m_{R}^{2})}{2\mu}\;$,
and $\: C_{x}(\tau)=Var_{x}e^{-\mu|\tau|}$.\\
\\
MATLAB was used to generate satisfaction profile of $U$, according
to the derived scheme of equations. This is done for a large number
of random combinations of $\mu$, number of favorable websites and
search duration (each randomly chosen).\\
\\
\textbf{4. Results}\\
With the minimum decay parameter (that is, a user with least irritation),
the satisfaction level is found to grow with both the duration of
search time and number of useful web-sites. Fig-1A to Fig-1C demonstrate
this trend with progressively increasing magnitude of maximum mark
of \textquoteleft{}Satisfaction index\textquoteright{} through the
stages (($\mu$:0.001, no. of websites:10, search duration:4000 seconds),
($\mu$:0.001, no. of websites:15, search duration:6000 seconds) and
($\mu$:0.001, no. of websites:20, search duration:9000 seconds)).
However, beyond this stage, the satisfaction profile attains saturation;
henceforth, the stage ($\mu$:0.001, no. of websites:30, search duration:12000
seconds) (Fig-1D) failed to show any marked improvement in user satisfaction
over the last stage described in Fig-1C. While this saturation mimics
convincingly the fatigue of a user after a long (3hour 20 minutes)
web-searching, the slow build-up of satisfaction profile (Fig-1B to
Fig-1D) reproduces the restless discontent of the user at the initial
stages of the search, before he identifies the correct strategy for
searching. With the increase in the magnitude of decay parameter $\mu\:$,
trend observed in Fig-1A to Fig-1D reverses noticeably. With an\textbf{
}increase in $\;\mu\:$, longer searches turn out to be less productive;
Fig-1E, 1F, and 1G demonstrate this trend clearly. It vindicates our
everyday experience. Even though $U$ comes across more number of
helpful web-sites during a longer search, higher magnitude of $\mu$(implying
the presence of increased level of frustration, irritation etc..)
results in user\textquoteright{}s missing out on the contents of helpful
web-sites. As a result, cumulative satisfaction profiles assume lower
magnitudes, progressively. With an even higher magnitude of $\mu\:$(describing
a dissatisfied user with significantly high level of irritation and
frustration), both the time for web-searching and the number of helpful
web-sites found, becomes irrelevant; as the cumulative satisfaction
profile struggles to reach even unity. This lays (theoretical) ground
for our experience that a web-search carried out with an enraged and
disgruntled mind can rarely present us with meaningful results.\\
\\
Thus, the general trend, somewhat unexpectedly\textbf{, }tends
to suggest the strongest dependence of user satisfaction level on
the decay parameter $\mu\:$, rather than on the number of favorable
web-sites. In other words, even if the user searches internet for
a long duration of time (say, for 3 continuous hours) and even if
he comes across a decent number of web-sites with helpful information
(say, 15, or in a lucky case, 20); his satisfaction level would still
be low, for a slightly high decay parameter (say, 0.008 or 0.01).
In the extreme cases, for a high magnitude of decay parameter (say
0.2, 0.5 etc.), cumulative satisfaction level will fail to reach the
positive mark at all. This is natural to expect; because, although
$\mu\:$ in the present model, is measured by non-availability of
pertinent information in web-sites, origin of $\mu\:$ may well be
rooted in the states of user's dis-satisfaction before he started
web-browsing. To summarize, present model predicts that all the efforts
of software engineers and designers to ensure satisfaction of the
users by providing pertinent information to them, will be meaningful
if only the users have their irritation and/or frustration at a low
level; that is within a limit $\:\mu<0.008$.\\
\\
Decay of the satisfaction profile is described in Fig-2, which
describes the steadiness(or the lack of it) of user satisfaction by
analyzing the correlations between satisfaction levels at random intervals
during the search operation. It implies that frustration of not finding
required information accumulates quickly on the previous states of
irritation, and hence, the cumulative satisfaction profile suffers
notably when $\mu\:$ undergoes a transition from 0.001 to 0.008,
or to 0.02. On the other hand, since the growth of satisfaction takes
place in a unit-step function (or multiples of it), a steady nature
in its cumulative profile could be observed. Since$\lambda$ is primarily
dependent on the content of the first page of a web-site, it can be
established that a web-site will gain much by presenting a glimpse
of all the information it can provide, on its very first page in simple
uncluttered form.\\
\\
To validate the proposed theory, results from three real-life
web-searching operations (Fig-3 suit) were compared to simulation
results presented in Fig-1 suit. Here, three users with least level
of initial irritation and different levels of web-browsing expertise
(comparatively naive, medium, and comparatively proficient) were asked
to search on three completely different topics; viz. cooking recipe,
destination for trekking and skiing, and research papers on possible
dependence of protein stability on the distribution of chiral centres
within it. Similarity of Fig-3 suit plots with the theoretically predicted
patterns (Fig-1 suit) validates the later. More importantly perhaps,
stationarity in user satisfaction profile predicted at the limiting
case of the theoretical model, matched perfectly with saturation of
satisfaction profile in two out of three case-studies. This implied
that the marker 'satisfaction retention quotient' could clearly distinguish
between the feeling of satisfaction out of comprehensive accumulation
of desired information (12/6980/0.001=1.719 for Fig-3A, 14/11500/0.001=1.217
for Fig-3B), and feeling of dis-satisfaction at the absence of desired
information (6/11360/0.001=0.528 for Fig-3C). Essence of typical web-searching
operations, therefore, could be objectively captured with the present
model.\\
\textbf{}\\
\textbf{}\\
\textbf{5. Discussions:}\\
Here we jot down the principal findings and inferences of the
present study.\\
\\
\textbf{5.1)} \textbf{Aggregation of fragmented micro information
:}\\
Rarely does a user receive the entire bulk of desired information
from one web-site. Instead, the comprehensive information is only
gathered by collecting and aggregating relevant pieces information
in quantized form, from various web-sites encountered during a particular
act of web navigation. Therefore, websites will gain more by, first,
stating unambiguously the scope and depth of information that they
can provide, preferably in the very first page; and second, by balancing
the contents of quantized information. The first of these might help
the sites to meet various visitors; who, even if not happy to know
the absence of desired information, will at least not leave the site
in an irritated state of mind. On the other hand, second condition
implies that a carefully planned balance between scope and depth of
the presented subject matter through sequence of pages, will help
users to maintain their focus. Since the web-sites can realistically
hope to provide a surfer with only a fragment of information he desires
to have, sites might benefit by presenting the largest scope in the
very first page, before building a series of (separate) pyramidal
peaks of heterogeneous depths upon it.\\
\\
\textbf{5.2) Capturing the attention of the user :}\\
Web-sites want to capture the attention of browsers. Present model
argues that the degree of user satisfaction depends entirely on the
degree of information received by the user during his visit to any
site. When a user feels favorable approach to some piece of information,
his satisfaction increases; when quanta of information fail to meet
visitor\textquoteright{}s expectation, his satisfaction decreases.
In the context of capturing the attention of a web-browser, role of
the first page of any website, plays a crucial role. The surfer will
continue to explore the website if only he finds this (first) page
to be rich in relevant quanta of information.\\
\\
\textbf{5.3) Measure of user satisfaction with the marker 'satisfaction
retention quotient':}\\
Construction of a suitable marker that describes evolution of
user satisfaction during web-browsing, is a difficult task. Simultaneous
evolution of three tangible processes, viz., growth of satisfaction
due to finding desired information, decay of satisfaction due to not
finding relevant information, and user's failure to remember every
piece of acquired information; alongside several other intangible
but influencing parameters (say, user's state of mind at the onset
of a particular web-search, etc..) - accounts for this difficulty.
To achieve precisely this (difficult) target, a marker named 'satisfaction
retention quotient' is proposed here. A magnitude of greater than
unity of this marker points to an instance of happy web-searching,
whereas a value less than unity implies a rather unhappy instance
of web searching. As a whole, while the markers proposed here ($\:\mu<0.008$
and 'satisfaction retention quotient') can potentially help engineers,
developers and user-interface designers associated with the emerging
field of Human-Computer-Interaction; objective description of growth
and decay of user's satisfaction might help practicing psychologists.
More importantly perhaps, we, the common Internet users, might find
this model to be useful, when every so often we wonder how happy or
unhappy our web-browsing was.\\
\\
\textbf{5.4) Growth (and decay) of satisfaction - is it linear
or non-linear?}\\
Common experiences tend to suggest that growth (or decay) of satisfaction
profile while searching the web has a non-linear nature. However,
the model proposed in the present work, although reasonably satisfactory,
is linear. - This situation seems paradoxical. Nonetheless, a careful
review of the various elements involved in web-searching process helps
us to resolve this paradox. One notes that since a web-searching operation
builds upon the serach query gradually, after the (possible) initial
lack of direction in the search query, the user learns to quickly
concentrate on subset of possible queries and attempts to focus on
it continuously. This process of continuous sharpening of the search
query, more often than not, results in a linearly additive accumulation
of satisfaction. Hence, although one might (loosely) perceive the
growth of satisfaction during web-searching to be non-linear, the
essence of the process remains linear. Thus, our assumption that satisfaction
of web-searching is primarily dependent on the content of the web-page
and not on other factors - stands vindicated.\\
\\
Nevertheless, it may assume importance to decipher the root of
(apparent) perception of non-linearity in evolution of satisfaction
profile. This demands an elaborate survey of various factors (at times
intangible) which may influence actual web-searching operation. To
concentrate on any one of these factors, we zoom in on the 'Site speed';
that is the speed of web-page loading in local browser. This is an
important parameter that influences satisfaction profile of any web-user
{[}21{]}, especially during an intense web-searching operation. One
may attribute the perceived non-linearity in growth (or decay) of
satisfaction to this factor (although many other intangible factors
may well be functional). Inclusion of this parameter in our model
will have undoubtedly resulted into an explicit time-dependent non-linear
model. However, the fact that the present model can ensure satisfactory
accuracy even without incorporating 'site speed' into consideration,
tends to suggest that either the disparities in site-loading speeds
have diminished to a significant extent in 2011 and most sites ensure
fast-loading; or, the users who carried out web-searching for the
validation of the present model were mature enough to not let site-loading
speeds influence their satisfaction growth (or decay) while searching
for a precise piece of information. Trivially, first possibility tends
to support our assumption of not taking 'site speed' into account
when constructing the model for the present problem. The second possibility,
however, demands a detailed investigation, because it depends more
on the psychological maturity (and state) of the user.\\
\\
Therefore, while the current model may serve as a template for
future works on similar lines, it is not complete. A probabilistic
model that incorporates 'site speed' and other tiny yet important
factors influencing web-user's psychology, may achieve that (elusive)
completeness.\\
\\
\textbf{}\\
\textbf{Acknowledgement :} This work was supported by COE-DBT(Department
of Biotechnology, Govt. of India) scheme.\\
\\
\textbf{\underbar{References :}}\\
{[}1{]} E. Lee, K. Kwon, D.Schumann, Segmenting the non-adopter
category in the diffusion of Internet banking, International Journ.
Bank Marketing, 23(2005): 414-437.\\
{[}2{]} V. Perumal, B. Shanmugam, Internet banking: boon or bane,
Journ. Internet Banking and Commerce, 9(2004):1-6.\\
{[}3{]} S. Rotchanakitumnuai, M. Speece, Corporate customer perspectives
on business value of Thai Internet banking, Journ. Electronic Commerce
Research, 5(2004): 270-286.\\
{[}4{]} G. Shergill, B. Li, Internet banking - An empirical investigation
of customers\textquoteright{} behaviour for online banking in New
Zealand, Journ. E-business, 5(2005):1-16.\\
{[}5{]} N. Siu, J. Mou, Measuring Service Quality in Internet
Banking: the Case of Hong Kong, Journ. International Consumer Marketing,
17(2005):97-114.\\
{[}6{]} K. Williamson, S. Lichtenstein, J. Sullivan, D. Schauder,
To Choose, or not to Choose: Exploring Australians' views about Internet
banking, International Journ. Technology and Human Interaction, 2(2006)17-33.\\
{[}7{]} B. Suh, I. Han, Effect of trust on customer acceptance
of Internet banking, Electronic Commerce Res. Appl. 1(2002)247-261.\\
{[}8{]} M. Durkin, In search of the Internet-banking customer:
exploring the use of decision styles, International Journ. Bank Marketing,
22(2004): 484-503.\\
{[}9{]} H. Li, C. Kuo, M. Russell, The impact of perceived channel
utilities, shopping orientations, and demographics on the consumer\textquoteright{}s
online buying behavior, Journ. Computer-Mediated Comm., 5(1999)0.\\
{[}10{]} P. Konana, S. Balasubramanian, Technology adoption and
usage as a social-psychological-economic phenomenon: A study of online
investing, Decision Support Systems, 39(2005)505-524.\\
{[}11{]} I. Kovalenko, N. Kuznetsov, V. Shurenkov,\\
$Models\; of\; random\; processes\::\; a\; handbook\; for\; mathematicians\; and\; engineers$
(CRC Press, Boca Raton, 1996).\\
{[}12{]} B. Gnedenko, $Theory\; of\; Probability\,,\,6th\; ed.$
(CRC Press, Boca Raton, 1998)\\
{[}13{]} S. M. Ross, $Stochastic\; Processes\:,\:2nd\; ed.$ (Wiley,
New York, 1996)\\
{[}14{]} S. M. Ross, $Introduction\; to\; probability\; models\,,\;9th\; ed.$
(Academic Press, San Diego, 2006)\\
{[}15{]} A. Y. Khinchin, $Mathematical\; methods\; in\; the\; theory\; of\; queeing\:(English\; translation)$,
(Charles Griffin, London, 1960).\\
{[}16{]} E. Wentzel, L. Ovcharov, $Applied\; Problems\; in\; Probability\; Theory\;(English\; translation)$,
(Mir, Moscow, 1986).\\
{[}17{]} W. Feller, $An\; introduction\; to\; probability\; theory\; and\; its\; application\,,\; vol-1\,,\:2nd\; ed.$,
(Wiley, New York, 1957).\\
{[}18{]} D. Cahoy, Fractional Poisson process in terms of alpha-stable
densities, PhD. Thesis, Case Western Reserve University, Cleveland,
Ohio, USA, 2007. \\
{[}19{]} D.R. Cox, V. Isham, $Point\; Processes$ (Chapman and
Hall, London, 1980)\\
{[}20{]} A. Mayerson, D. Jones, N. Bowers, On the credibility
of the pure premium, Proc. Casualty Actuarial Soc., 55 (1968) 175\textendash{}185\textbf{.}\\
{[}21{]} Constantinides E., Influencing the online consumer\textquoteright{}s
behavior: the Web experience, Internet Research, 14(2004): 111-126.\\
\\
\\
\\
\textbf{Figure titles and legends :}\\
\textbf{Fig-1 suit)}: Title : User Satisfaction as function of
duration of searching time, number of favorable web-sites and decay
parameter (frustration level). \textbf{Fig-1A )} : Due to encountering
10 favorable web-sites in only 4000 seconds, the satisfaction index
grows rapidly. However, it could merely reach 1800 mark, due to relatively
low searching-time. \textbf{Fig-1B} ) : Slow-build up of cumulative
satisfaction, ultimately crossing 2000 mark; vindicating that it is
more probable for a search to come across favorable web-sites if it
is carried out on longer time interval. \textbf{Fig-1C} confirms this.
\textbf{Fig-1D} ) : Demonstrates that growth of satisfaction over
time does not follow a linear relationship; even though such a search
operation might increase the probability of coming across number of
helpful web-sites. It shows saturation in the cumulative profile of
satisfaction index. \textbf{Fig-1E} ) : With the increase in the magnitude
of $\mu$, trend observed in Fig-1A to Fig-1D reverses. With a slightly
higher $\mu$, longer searches turn out to be less productive, satisfaction
index decreases gradually. Fig-1F, Supp.Fig-1C to Supp.Fig-1E confirm
this. \textbf{Fig-1F} ) : With an even higher magnitude of $\mu$(describing
a dissatisfied user with significantly high level of irritation and
frustration), the time for web-searching and the number if helpful
web-sites found, becomes rather irrelevant; the cumulative satisfaction
profile struggles to reach unity. This vindicates our common experience
that a web-search carried out with an enraged and disgruntled mind
can rarely present us with meaningful results. \textbf{Fig-1G}, \textbf{Fig-1H}
and Supp.Fig-1F to Supp.Fig-1I confirm this. \textbf{Fig-2} ) : Title
: Correlation between satisfaction levels. Describes how the steadiness
of cumulative satisfaction level suffers from increasing magnitude
of decay parameter. \textbf{Fig-3 suit)}: Title: Validation of theoretically
predicted scheme. \textbf{Fig-3A} ): Title : Search for a particular
cooking recipe. Satisfaction levels as reported by a (relatively inexperienced)
Internet user while searching for the recipe of typical Indian dish.
\textbf{Fig-3B} ): Title : Search for a particular trekking and skiing
destination. Satisfaction levels as reported by an experienced Internet
user while searching for a trekking and skiing destination, somewhere
in Himalaya. \textbf{Fig-3C} ): Title : Search for scientific literature
on a specialized topic. Satisfaction levels as reported by a highly
experienced Internet user while searching through research papers
for information on possible dependence of protein stability on the
distribution of chiral centres within it.\\
\\
\\

\end{document}